\documentstyle[12pt]{article}
\title{$q$-Calculus Framework For Entropy In Multifractal Distributions}
\author{{Ramandeep S. Johal,}\footnote{Postal Address: 1110, 36-C,  
Chandigarh -160014, India} \\
{\it Department of Physics, Panjab University,} \\
{\it Chandigarh -166014, India. }\\e-mail: raman\%phys@puniv.chd.nic.in} 
\begin{document}
\baselineskip 24pt
\date{}
\maketitle
\def\be{\begin{equation}}
\def\ee{\end{equation}}
\def\ba{\begin{eqnarray}}
\def\ea{\end{eqnarray}}
\def\al{$\alpha _i$}
\def\qder{$\Cal D _{q,\alpha _i}$}
\begin{abstract}
The connection between Tsallis entropy for a multifractal distribution  
and Jackson's $q$-derivative is established.
Based on this derivation and definition of a homogeneous function, a 
$q$-analogue of Shannon's entropy is
discussed. $q$-additivity of this entropy is shown. We also define 
$q$-analogue of Kullback relative entropy. The
implications of lattice structure beneath $q$-calculus are highlighted  
in the context of $q$-entropy.
\end{abstract}
Non-extensive Tsallis Thermostatistics ( NTT) \cite{rf1} 
generalizes the Boltzmann-Gibbs (BG) statistics, to treat the
non-extensivity of physical systems \cite{rf2}. It has been 
applied with success to many different situations (for complete
reviews, see ref. \cite{rf3}). One non-extensive quantity which
is playing a useful role  
is Tsallis entropy \cite{rf4}. Given a probability distribution
$\{p_i\}_{i=1,...,W}$ where $i$ is the index for system configuration, 
Tsallis entropy is given by
\be
S^{T}_{q} = \frac{1-\sum_{i=1}^{W} (p_i)^q}{q-1}. \label{en1}
\ee
$q$ is a real parameter, assumed to be positive. $W$ is the number of  
accessible configurations.  Boltzmann's constant
$k_B$ has been set equal to unity. As $q\to 1$, $S^{T}_{q} \to 
-\sum p_i {\rm ln}\;p_i$, which is Shannon entropy. Thus parameter $q$
describes deviations of Tsallis entropy from Shannon entropy.

On a different side, quantum algebras and in general, $q$-deformed physical
theories \cite{rf5} have been subject of great attention in the last 
decade. An important feature of these theories is the presence of one
(or more) deformation parameter $q$, which describes deviation from
standard Lie symmetries. Usually, for $q\to 1$, the formalism reverts 
to the standard one.

Recently, Tsallis noted \cite{rf4} a similarity between $q$-numbers used in 
$q$-deformed theories and the entropy of eq. (\ref{en1}).  Notably the  
pseudo-additive property of 
both quantities is alike. Abe \cite{rf6} provided a more analytic 
foundation to this connection, based on $q$-calculus.  Also in \cite{rf7},  
$q$-deformed Landau diamagntism was studied to strengthen the 
relation between qunatum groups and NTT.
 
Originally, Tsallis' proposal was aimed to accomodate scale invariance 
in a system with multifractal properties to the thermodynamic formalism. 
In this communication,  we
clarify the relation between $q$-calculus and multifractal properties of 
a probability 
distribution. We also propose a more general defintion  of $q$-entropy 
based on $q$-calculus.
It is shown that pseudo-additivity of this entropy folows from 
$q$-additivity of $q$-numbers.
Moreover, Tsallis entropy can also be accomodated in this definition. 
The lattice structure behind the $q$-calculus framework is also highlighted.

For a multifractal distribution,  we assume a local scaling for the 
probabilites,
\be
p_i = R^{\alpha _i} \label{en2},
\ee
where $\{\alpha _i\}_{i=1,...,W}$ is the set of scaling indices. 
$R$ is the size of each box used to cover 
the phase space.  Understandably, $p_i$ can be interpreted as the 
probability of visiting the
box labelled with $i$ and having a scaling index  $\alpha _i$. 
Thus, in a manner similar to 
suggested in
\cite{rf6}, we can write Shannon's entropy as
\be
-\sum_{i}\alpha _i \frac{d}{d\alpha _i} p_i =  -\sum _{i} p_i 
{\rm ln}\;p_i.\label{en3}
\ee
Note however, that the variable $\alpha _i$ here can be given a suitable 
interpretation  which was
a dummy variable in \cite{rf6}.

If we replace the ordianry derivative in (\ref{en3}) with jackson's 
$q$-derivative \cite{rf8}, we get
\be
-\sum_{i} \alpha _i {{\cal D}^{q}_{\alpha _i}} p_i = \frac{1-\sum_{i=1}^{W} 
(p_i)^q}{q-1}, 
\label{en4}
\ee
which is Tsallis entropy. Instead of jackson's derivative, if we use the 
symmetric  $q$-derivative
which has $q\leftrightarrow q^{-1}$ invariance, we obtain the alternate 
entropy suggested in 
\cite{rf6}. In the following, we concentrate on the entropy based on 
Jackson's derivative.

We argue that although use of ordinary derivative w.r.t. $\alpha _i$ in 
(\ref{en3}) is 
mathematically correct, it is not proper in an {\it operational} sense. 
Note that even though the 
size of the boxes is taken to be small, both the size and their number is 
finite.  So it looks more 
reasonable to use $q$-derivative which involves dilatation of the argument 
$\alpha _i$ than the 
ordinary derivative which takes into account infinitsimal changes of argument. 

Consider now a generalized probability distribution $\{p_i\}$ where 
$p_i(\alpha _i)$ is 
homogeneous function of degree $a_i$ and $\alpha _i$ is not necessarily 
a scaling index. Then 
by definition
\be
\alpha _i {{\cal D}^{q}_{\alpha _i}}  p_i(\alpha _i) = [a_i] p_i(\alpha _i), 
\label{en5}
\ee
where $[a_i] = \frac{q^{a_i}-1}{q-1}$ is the Jackson $q$-number. Then we 
define the $q$-entropy as
\be
-\sum_{i} \alpha _i {{\cal D}^{q}_{\alpha _i}}  p_i(\alpha _i) = -\sum_{i} 
[a_i] p_i(\alpha _i). 
\label{en6}
\ee
As $q\to 1$, we get
\be
-\sum_{i}\alpha _i \frac{d}{d\alpha _i} p_i =  -\sum _{i} a_ip_i. \label{rf7}
\ee
If we identify, $a_i ={\rm ln}\;p_i$ in (\ref{rf7}), we get Shannon entropy 
on the r.h.s. . Alternatively, if we set 
$\alpha _i$ as the local scaling index,  we again obtain Shannon entropy 
as defined in 
(\ref{en3}). Similarly, if we take $\alpha _i$ in (\ref{en6}) as scaling 
index, then equality of 
(\ref{en4}) and (\ref{en6}) gives
\be
[a_i] = \frac{q^{a_i}-1}{q-1} = \frac{(p_i)^{q-1}-1}{q-1} \label{en8}
\ee
which gives
\be
a_i = \frac{q-1}{{\rm ln}\;q} {\rm ln}\;p_i \label{en9}
\ee
Thus we can alternatively define Tsallis entropy as {\it the negative of mean
 of $[a_i]$'s over the probability distribution}, where $a_i$ is given 
by (\ref{en9}).

Tsallis entropy can be looked upon as a $q$-deformed Shannon entropy \cite{rf6}.
Thus by setting $a_i={\rm ln}\;p_i$ in the definition of $q$-entropy (in 
order to 
obtain Shannon entropy in the limit $q\to 1$, we get another $q$-deformed 
analogue
of Shannon entropy
\be
{S_q}^{\prime} = -\sum_{i} [{\rm ln}\;p_i] p_i. \label{en10}
\ee
Note that ${S_q}^{\prime}>S^{T}_{q}$ for $0<q<1$.  In fig. 1 we compare 
the $i$-th term of 
Shannon, Tsallis and $q$-entropy by taking $p_i=0.2$. Only entropy values 
for $q<1$  appear
to be physically meaningful, as discussed below. Thus ${S_q}^{\prime}$  
is a concave function
due to a similar property of Tsallis entropy. 

To see the additive property of ${S_q}^{\prime}$, consider two independent 
subsystems $I$ and $II$, described
by normalized probability distributions $\{p_i\}$ and $\{p_j\}$ respectively.
Then the $q$-entropy of the combined system may be written as
\ba
{S_q}^{\prime}(I+II) & = & -\sum_{i,j} [{\rm ln}\;p_{ij}] p_{ij},\nonumber \\
                     & = & -\sum_{i,j}[{\rm ln}\;p_i + {\rm ln}\;p_j]p_ip_j 
                              \nonumber \\
                     &  =& {S_q}^{\prime}(I) + {S_q}^{\prime}(II)
                         + (1-q){S_q}^{\prime}(I){S_q}^{\prime}(II). \ea
We have made use of the pseudo-additive property of the $q$-numbers and the 
normalization 
property of the probability distributions. The similar property of Tsallis 
entropy can be seen to emerge because of the relation (\ref{en9}).

We may also define the $q$-analogue of Kullback relative entropy \cite{rf10}, 
in going from a probability distribution $p^0$ to another one $p$.
Consider the difference $[a_i]-[a^{0}_{i}]$, where $a_i={\rm ln}\;p_i$ and
$a^{0}_{i}= {\rm ln}\;p^{0}_{i}$. The average weighted against the new 
probability distribution, gives the $q$-analogue of the Kullback 
relative entropy,
\be
K_q(p,p^0) = \sum_ip_i([a_i]-[a^{0}_{i}]).
\ee
In the limit of $q\to 1$, we get the standard Kullback relative entropy.
Using $K_q(p^0,p) = \sum_ip^{0}_i([a^{0}_{i}]-[a_{i}])$, we obtain the 
$q$-analogue of the symmetric sum,
\ba
D_q(p,p^0) & = & K_q(p,p^0) + K_q(p^0,p)\nonumber \\
           & = & \sum_{i} ([a_i]-[a^{0}_{i}])(p_i-p^{0}_{i})
\ea
Each term in the sum is positive and is zero iff $p_i= p^{0}_{i}$. Thus
this function appears suitable for a metric in the functional space of
probability distributions.

Finally, we remark on the lattice structure which underlies the $q$-calculus 
framework of entropy. A natural lattice already exists, because we 
partition the
phase space into boxes of equal size $R$. The lattice constant $R$ can be 
identified
with $|q-1|$. Thus $q\to 1$ limit also implies $R\to 0$. The finite size 
of the 
boxes causes coarse graining of the phase space, as a result the information
we would have about the structure of the distribution is also coarse grained. 
Thus
the value of generalized $q$-entropies should be greater than the  Shannon 
entropy,
which is shown here as the limit of $q\to 1$ case. We note that both Tsallis
and $q$-entropy satisfy this condition for $q<1$ (Fig. 1). The divergence of 
$q$-entropy
can also be explained because as $q\to 0$, the size of the boxes increases, 
which
causes more loss of information and thus increase in entropy.

The author would like to thank Sumiyoshi Abe for encouraging remarks and
useful discussions.

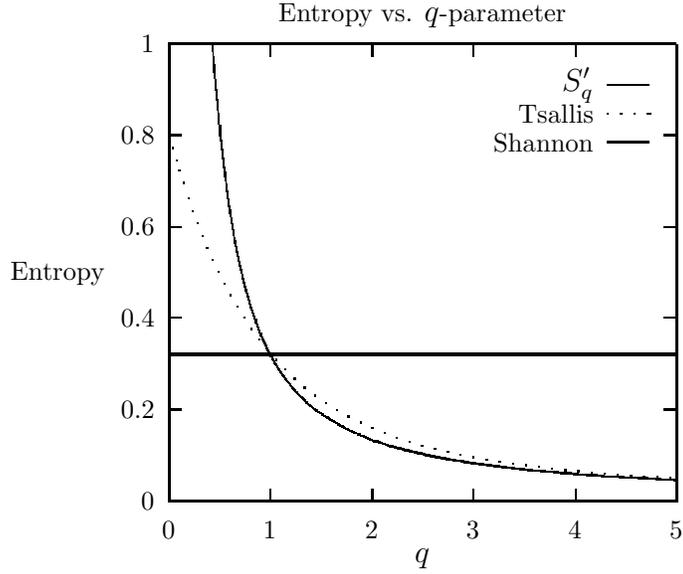
\begin{figure}
\caption{ Fig.1 : Plots of $i$-th term of  entropies for $p_i=0.2$ against 
the deformation parameter $q$.}
\setlength{\unitlength}{0.240900pt}
\ifx\plotpoint\undefined\newsavebox{\plotpoint}\fi
\sbox{\plotpoint}{\rule[-0.200pt]{0.400pt}{0.400pt}}%
\begin{picture}(1081,900)(0,0)
\font\gnuplot=cmr10 at 10pt
\gnuplot
\sbox{\plotpoint}{\rule[-0.200pt]{0.400pt}{0.400pt}}%
\put(220.0,113.0){\rule[-0.200pt]{191.997pt}{0.400pt}}
\put(220.0,113.0){\rule[-0.200pt]{0.400pt}{173.207pt}}
\put(220.0,113.0){\rule[-0.200pt]{4.818pt}{0.400pt}}
\put(198,113){\makebox(0,0)[r]{0}}
\put(997.0,113.0){\rule[-0.200pt]{4.818pt}{0.400pt}}
\put(220.0,257.0){\rule[-0.200pt]{4.818pt}{0.400pt}}
\put(198,257){\makebox(0,0)[r]{0.2}}
\put(997.0,257.0){\rule[-0.200pt]{4.818pt}{0.400pt}}
\put(220.0,401.0){\rule[-0.200pt]{4.818pt}{0.400pt}}
\put(198,401){\makebox(0,0)[r]{0.4}}
\put(997.0,401.0){\rule[-0.200pt]{4.818pt}{0.400pt}}
\put(220.0,544.0){\rule[-0.200pt]{4.818pt}{0.400pt}}
\put(198,544){\makebox(0,0)[r]{0.6}}
\put(997.0,544.0){\rule[-0.200pt]{4.818pt}{0.400pt}}
\put(220.0,688.0){\rule[-0.200pt]{4.818pt}{0.400pt}}
\put(198,688){\makebox(0,0)[r]{0.8}}
\put(997.0,688.0){\rule[-0.200pt]{4.818pt}{0.400pt}}
\put(220.0,832.0){\rule[-0.200pt]{4.818pt}{0.400pt}}
\put(198,832){\makebox(0,0)[r]{1}}
\put(997.0,832.0){\rule[-0.200pt]{4.818pt}{0.400pt}}
\put(220.0,113.0){\rule[-0.200pt]{0.400pt}{4.818pt}}
\put(220,68){\makebox(0,0){0}}
\put(220.0,812.0){\rule[-0.200pt]{0.400pt}{4.818pt}}
\put(379.0,113.0){\rule[-0.200pt]{0.400pt}{4.818pt}}
\put(379,68){\makebox(0,0){1}}
\put(379.0,812.0){\rule[-0.200pt]{0.400pt}{4.818pt}}
\put(539.0,113.0){\rule[-0.200pt]{0.400pt}{4.818pt}}
\put(539,68){\makebox(0,0){2}}
\put(539.0,812.0){\rule[-0.200pt]{0.400pt}{4.818pt}}
\put(698.0,113.0){\rule[-0.200pt]{0.400pt}{4.818pt}}
\put(698,68){\makebox(0,0){3}}
\put(698.0,812.0){\rule[-0.200pt]{0.400pt}{4.818pt}}
\put(858.0,113.0){\rule[-0.200pt]{0.400pt}{4.818pt}}
\put(858,68){\makebox(0,0){4}}
\put(858.0,812.0){\rule[-0.200pt]{0.400pt}{4.818pt}}
\put(1017.0,113.0){\rule[-0.200pt]{0.400pt}{4.818pt}}
\put(1017,68){\makebox(0,0){5}}
\put(1017.0,812.0){\rule[-0.200pt]{0.400pt}{4.818pt}}
\put(220.0,113.0){\rule[-0.200pt]{191.997pt}{0.400pt}}
\put(1017.0,113.0){\rule[-0.200pt]{0.400pt}{173.207pt}}
\put(220.0,832.0){\rule[-0.200pt]{191.997pt}{0.400pt}}
\put(45,472){\makebox(0,0){Entropy}}
\put(618,23){\makebox(0,0){$q$}}
\put(618,877){\makebox(0,0){Entropy vs. $q$-parameter}}
\put(220.0,113.0){\rule[-0.200pt]{0.400pt}{173.207pt}}
\put(887,767){\makebox(0,0)[r]{$S_q^{\prime}$}}
\multiput(289.61,806.68)(0.447,-9.839){3}{\rule{0.108pt}{6.100pt}}
\multiput(288.17,819.34)(3.000,-32.339){2}{\rule{0.400pt}{3.050pt}}
\multiput(292.59,769.61)(0.489,-5.329){15}{\rule{0.118pt}{4.189pt}}
\multiput(291.17,778.31)(9.000,-83.306){2}{\rule{0.400pt}{2.094pt}}
\multiput(301.59,679.64)(0.488,-4.720){13}{\rule{0.117pt}{3.700pt}}
\multiput(300.17,687.32)(8.000,-64.320){2}{\rule{0.400pt}{1.850pt}}
\multiput(309.59,610.55)(0.488,-3.796){13}{\rule{0.117pt}{3.000pt}}
\multiput(308.17,616.77)(8.000,-51.773){2}{\rule{0.400pt}{1.500pt}}
\multiput(317.59,554.83)(0.488,-3.069){13}{\rule{0.117pt}{2.450pt}}
\multiput(316.17,559.91)(8.000,-41.915){2}{\rule{0.400pt}{1.225pt}}
\multiput(325.59,509.70)(0.488,-2.475){13}{\rule{0.117pt}{2.000pt}}
\multiput(324.17,513.85)(8.000,-33.849){2}{\rule{0.400pt}{1.000pt}}
\multiput(333.59,472.74)(0.488,-2.145){13}{\rule{0.117pt}{1.750pt}}
\multiput(332.17,476.37)(8.000,-29.368){2}{\rule{0.400pt}{0.875pt}}
\multiput(341.59,440.98)(0.488,-1.748){13}{\rule{0.117pt}{1.450pt}}
\multiput(340.17,443.99)(8.000,-23.990){2}{\rule{0.400pt}{0.725pt}}
\multiput(349.59,414.60)(0.488,-1.550){13}{\rule{0.117pt}{1.300pt}}
\multiput(348.17,417.30)(8.000,-21.302){2}{\rule{0.400pt}{0.650pt}}
\multiput(357.59,391.23)(0.488,-1.352){13}{\rule{0.117pt}{1.150pt}}
\multiput(356.17,393.61)(8.000,-18.613){2}{\rule{0.400pt}{0.575pt}}
\multiput(365.59,370.85)(0.488,-1.154){13}{\rule{0.117pt}{1.000pt}}
\multiput(364.17,372.92)(8.000,-15.924){2}{\rule{0.400pt}{0.500pt}}
\multiput(373.59,353.26)(0.488,-1.022){13}{\rule{0.117pt}{0.900pt}}
\multiput(372.17,355.13)(8.000,-14.132){2}{\rule{0.400pt}{0.450pt}}
\multiput(381.59,337.68)(0.488,-0.890){13}{\rule{0.117pt}{0.800pt}}
\multiput(380.17,339.34)(8.000,-12.340){2}{\rule{0.400pt}{0.400pt}}
\multiput(389.59,324.09)(0.488,-0.758){13}{\rule{0.117pt}{0.700pt}}
\multiput(388.17,325.55)(8.000,-10.547){2}{\rule{0.400pt}{0.350pt}}
\multiput(397.59,312.30)(0.488,-0.692){13}{\rule{0.117pt}{0.650pt}}
\multiput(396.17,313.65)(8.000,-9.651){2}{\rule{0.400pt}{0.325pt}}
\multiput(405.59,301.51)(0.488,-0.626){13}{\rule{0.117pt}{0.600pt}}
\multiput(404.17,302.75)(8.000,-8.755){2}{\rule{0.400pt}{0.300pt}}
\multiput(413.59,291.51)(0.488,-0.626){13}{\rule{0.117pt}{0.600pt}}
\multiput(412.17,292.75)(8.000,-8.755){2}{\rule{0.400pt}{0.300pt}}
\multiput(421.00,282.93)(0.494,-0.488){13}{\rule{0.500pt}{0.117pt}}
\multiput(421.00,283.17)(6.962,-8.000){2}{\rule{0.250pt}{0.400pt}}
\multiput(429.00,274.93)(0.494,-0.488){13}{\rule{0.500pt}{0.117pt}}
\multiput(429.00,275.17)(6.962,-8.000){2}{\rule{0.250pt}{0.400pt}}
\multiput(437.00,266.93)(0.569,-0.485){11}{\rule{0.557pt}{0.117pt}}
\multiput(437.00,267.17)(6.844,-7.000){2}{\rule{0.279pt}{0.400pt}}
\multiput(445.00,259.93)(0.671,-0.482){9}{\rule{0.633pt}{0.116pt}}
\multiput(445.00,260.17)(6.685,-6.000){2}{\rule{0.317pt}{0.400pt}}
\multiput(453.00,253.93)(0.762,-0.482){9}{\rule{0.700pt}{0.116pt}}
\multiput(453.00,254.17)(7.547,-6.000){2}{\rule{0.350pt}{0.400pt}}
\multiput(462.00,247.93)(0.821,-0.477){7}{\rule{0.740pt}{0.115pt}}
\multiput(462.00,248.17)(6.464,-5.000){2}{\rule{0.370pt}{0.400pt}}
\multiput(470.00,242.93)(0.821,-0.477){7}{\rule{0.740pt}{0.115pt}}
\multiput(470.00,243.17)(6.464,-5.000){2}{\rule{0.370pt}{0.400pt}}
\multiput(478.00,237.93)(0.821,-0.477){7}{\rule{0.740pt}{0.115pt}}
\multiput(478.00,238.17)(6.464,-5.000){2}{\rule{0.370pt}{0.400pt}}
\multiput(486.00,232.94)(1.066,-0.468){5}{\rule{0.900pt}{0.113pt}}
\multiput(486.00,233.17)(6.132,-4.000){2}{\rule{0.450pt}{0.400pt}}
\multiput(494.00,228.93)(0.821,-0.477){7}{\rule{0.740pt}{0.115pt}}
\multiput(494.00,229.17)(6.464,-5.000){2}{\rule{0.370pt}{0.400pt}}
\multiput(502.00,223.95)(1.579,-0.447){3}{\rule{1.167pt}{0.108pt}}
\multiput(502.00,224.17)(5.579,-3.000){2}{\rule{0.583pt}{0.400pt}}
\multiput(510.00,220.94)(1.066,-0.468){5}{\rule{0.900pt}{0.113pt}}
\multiput(510.00,221.17)(6.132,-4.000){2}{\rule{0.450pt}{0.400pt}}
\multiput(518.00,216.95)(1.579,-0.447){3}{\rule{1.167pt}{0.108pt}}
\multiput(518.00,217.17)(5.579,-3.000){2}{\rule{0.583pt}{0.400pt}}
\multiput(526.00,213.94)(1.066,-0.468){5}{\rule{0.900pt}{0.113pt}}
\multiput(526.00,214.17)(6.132,-4.000){2}{\rule{0.450pt}{0.400pt}}
\multiput(534.00,209.95)(1.579,-0.447){3}{\rule{1.167pt}{0.108pt}}
\multiput(534.00,210.17)(5.579,-3.000){2}{\rule{0.583pt}{0.400pt}}
\put(542,206.17){\rule{1.700pt}{0.400pt}}
\multiput(542.00,207.17)(4.472,-2.000){2}{\rule{0.850pt}{0.400pt}}
\multiput(550.00,204.95)(1.579,-0.447){3}{\rule{1.167pt}{0.108pt}}
\multiput(550.00,205.17)(5.579,-3.000){2}{\rule{0.583pt}{0.400pt}}
\multiput(558.00,201.95)(1.579,-0.447){3}{\rule{1.167pt}{0.108pt}}
\multiput(558.00,202.17)(5.579,-3.000){2}{\rule{0.583pt}{0.400pt}}
\put(566,198.17){\rule{1.700pt}{0.400pt}}
\multiput(566.00,199.17)(4.472,-2.000){2}{\rule{0.850pt}{0.400pt}}
\put(574,196.17){\rule{1.700pt}{0.400pt}}
\multiput(574.00,197.17)(4.472,-2.000){2}{\rule{0.850pt}{0.400pt}}
\put(582,194.17){\rule{1.700pt}{0.400pt}}
\multiput(582.00,195.17)(4.472,-2.000){2}{\rule{0.850pt}{0.400pt}}
\put(590,192.17){\rule{1.700pt}{0.400pt}}
\multiput(590.00,193.17)(4.472,-2.000){2}{\rule{0.850pt}{0.400pt}}
\put(598,190.17){\rule{1.700pt}{0.400pt}}
\multiput(598.00,191.17)(4.472,-2.000){2}{\rule{0.850pt}{0.400pt}}
\put(606,188.17){\rule{1.700pt}{0.400pt}}
\multiput(606.00,189.17)(4.472,-2.000){2}{\rule{0.850pt}{0.400pt}}
\put(614,186.17){\rule{1.900pt}{0.400pt}}
\multiput(614.00,187.17)(5.056,-2.000){2}{\rule{0.950pt}{0.400pt}}
\put(623,184.17){\rule{1.700pt}{0.400pt}}
\multiput(623.00,185.17)(4.472,-2.000){2}{\rule{0.850pt}{0.400pt}}
\put(631,182.67){\rule{1.927pt}{0.400pt}}
\multiput(631.00,183.17)(4.000,-1.000){2}{\rule{0.964pt}{0.400pt}}
\put(639,181.17){\rule{1.700pt}{0.400pt}}
\multiput(639.00,182.17)(4.472,-2.000){2}{\rule{0.850pt}{0.400pt}}
\put(647,179.67){\rule{1.927pt}{0.400pt}}
\multiput(647.00,180.17)(4.000,-1.000){2}{\rule{0.964pt}{0.400pt}}
\put(655,178.17){\rule{1.700pt}{0.400pt}}
\multiput(655.00,179.17)(4.472,-2.000){2}{\rule{0.850pt}{0.400pt}}
\put(663,176.67){\rule{1.927pt}{0.400pt}}
\multiput(663.00,177.17)(4.000,-1.000){2}{\rule{0.964pt}{0.400pt}}
\put(671,175.67){\rule{1.927pt}{0.400pt}}
\multiput(671.00,176.17)(4.000,-1.000){2}{\rule{0.964pt}{0.400pt}}
\put(679,174.17){\rule{1.700pt}{0.400pt}}
\multiput(679.00,175.17)(4.472,-2.000){2}{\rule{0.850pt}{0.400pt}}
\put(687,172.67){\rule{1.927pt}{0.400pt}}
\multiput(687.00,173.17)(4.000,-1.000){2}{\rule{0.964pt}{0.400pt}}
\put(695,171.67){\rule{1.927pt}{0.400pt}}
\multiput(695.00,172.17)(4.000,-1.000){2}{\rule{0.964pt}{0.400pt}}
\put(703,170.67){\rule{1.927pt}{0.400pt}}
\multiput(703.00,171.17)(4.000,-1.000){2}{\rule{0.964pt}{0.400pt}}
\put(711,169.67){\rule{1.927pt}{0.400pt}}
\multiput(711.00,170.17)(4.000,-1.000){2}{\rule{0.964pt}{0.400pt}}
\put(719,168.67){\rule{1.927pt}{0.400pt}}
\multiput(719.00,169.17)(4.000,-1.000){2}{\rule{0.964pt}{0.400pt}}
\put(727,167.67){\rule{1.927pt}{0.400pt}}
\multiput(727.00,168.17)(4.000,-1.000){2}{\rule{0.964pt}{0.400pt}}
\put(735,166.67){\rule{1.927pt}{0.400pt}}
\multiput(735.00,167.17)(4.000,-1.000){2}{\rule{0.964pt}{0.400pt}}
\put(743,165.67){\rule{1.927pt}{0.400pt}}
\multiput(743.00,166.17)(4.000,-1.000){2}{\rule{0.964pt}{0.400pt}}
\put(751,164.67){\rule{1.927pt}{0.400pt}}
\multiput(751.00,165.17)(4.000,-1.000){2}{\rule{0.964pt}{0.400pt}}
\put(759,163.67){\rule{1.927pt}{0.400pt}}
\multiput(759.00,164.17)(4.000,-1.000){2}{\rule{0.964pt}{0.400pt}}
\put(767,162.67){\rule{1.927pt}{0.400pt}}
\multiput(767.00,163.17)(4.000,-1.000){2}{\rule{0.964pt}{0.400pt}}
\put(775,161.67){\rule{2.168pt}{0.400pt}}
\multiput(775.00,162.17)(4.500,-1.000){2}{\rule{1.084pt}{0.400pt}}
\put(784,160.67){\rule{1.927pt}{0.400pt}}
\multiput(784.00,161.17)(4.000,-1.000){2}{\rule{0.964pt}{0.400pt}}
\put(909.0,767.0){\rule[-0.200pt]{15.899pt}{0.400pt}}
\put(800,159.67){\rule{1.927pt}{0.400pt}}
\multiput(800.00,160.17)(4.000,-1.000){2}{\rule{0.964pt}{0.400pt}}
\put(808,158.67){\rule{1.927pt}{0.400pt}}
\multiput(808.00,159.17)(4.000,-1.000){2}{\rule{0.964pt}{0.400pt}}
\put(792.0,161.0){\rule[-0.200pt]{1.927pt}{0.400pt}}
\put(824,157.67){\rule{1.927pt}{0.400pt}}
\multiput(824.00,158.17)(4.000,-1.000){2}{\rule{0.964pt}{0.400pt}}
\put(832,156.67){\rule{1.927pt}{0.400pt}}
\multiput(832.00,157.17)(4.000,-1.000){2}{\rule{0.964pt}{0.400pt}}
\put(816.0,159.0){\rule[-0.200pt]{1.927pt}{0.400pt}}
\put(848,155.67){\rule{1.927pt}{0.400pt}}
\multiput(848.00,156.17)(4.000,-1.000){2}{\rule{0.964pt}{0.400pt}}
\put(856,154.67){\rule{1.927pt}{0.400pt}}
\multiput(856.00,155.17)(4.000,-1.000){2}{\rule{0.964pt}{0.400pt}}
\put(840.0,157.0){\rule[-0.200pt]{1.927pt}{0.400pt}}
\put(872,153.67){\rule{1.927pt}{0.400pt}}
\multiput(872.00,154.17)(4.000,-1.000){2}{\rule{0.964pt}{0.400pt}}
\put(864.0,155.0){\rule[-0.200pt]{1.927pt}{0.400pt}}
\put(888,152.67){\rule{1.927pt}{0.400pt}}
\multiput(888.00,153.17)(4.000,-1.000){2}{\rule{0.964pt}{0.400pt}}
\put(896,151.67){\rule{1.927pt}{0.400pt}}
\multiput(896.00,152.17)(4.000,-1.000){2}{\rule{0.964pt}{0.400pt}}
\put(880.0,154.0){\rule[-0.200pt]{1.927pt}{0.400pt}}
\put(912,150.67){\rule{1.927pt}{0.400pt}}
\multiput(912.00,151.17)(4.000,-1.000){2}{\rule{0.964pt}{0.400pt}}
\put(904.0,152.0){\rule[-0.200pt]{1.927pt}{0.400pt}}
\put(928,149.67){\rule{1.927pt}{0.400pt}}
\multiput(928.00,150.17)(4.000,-1.000){2}{\rule{0.964pt}{0.400pt}}
\put(920.0,151.0){\rule[-0.200pt]{1.927pt}{0.400pt}}
\put(953,148.67){\rule{1.927pt}{0.400pt}}
\multiput(953.00,149.17)(4.000,-1.000){2}{\rule{0.964pt}{0.400pt}}
\put(936.0,150.0){\rule[-0.200pt]{4.095pt}{0.400pt}}
\put(969,147.67){\rule{1.927pt}{0.400pt}}
\multiput(969.00,148.17)(4.000,-1.000){2}{\rule{0.964pt}{0.400pt}}
\put(961.0,149.0){\rule[-0.200pt]{1.927pt}{0.400pt}}
\put(985,146.67){\rule{1.927pt}{0.400pt}}
\multiput(985.00,147.17)(4.000,-1.000){2}{\rule{0.964pt}{0.400pt}}
\put(977.0,148.0){\rule[-0.200pt]{1.927pt}{0.400pt}}
\put(1009,145.67){\rule{1.927pt}{0.400pt}}
\multiput(1009.00,146.17)(4.000,-1.000){2}{\rule{0.964pt}{0.400pt}}
\put(993.0,147.0){\rule[-0.200pt]{3.854pt}{0.400pt}}
\put(887,722){\makebox(0,0)[r]{Tsallis}}
\multiput(909,722)(20.756,0.000){4}{\usebox{\plotpoint}}
\put(975,722){\usebox{\plotpoint}}
\put(220,688){\usebox{\plotpoint}}
\multiput(220,688)(5.702,-19.957){2}{\usebox{\plotpoint}}
\put(231.52,648.12){\usebox{\plotpoint}}
\multiput(236,633)(6.326,-19.768){2}{\usebox{\plotpoint}}
\put(250.40,588.79){\usebox{\plotpoint}}
\put(257.16,569.17){\usebox{\plotpoint}}
\put(264.31,549.68){\usebox{\plotpoint}}
\put(271.86,530.35){\usebox{\plotpoint}}
\put(279.73,511.15){\usebox{\plotpoint}}
\put(288.15,492.18){\usebox{\plotpoint}}
\put(297.48,473.65){\usebox{\plotpoint}}
\put(306.92,455.16){\usebox{\plotpoint}}
\put(316.58,436.79){\usebox{\plotpoint}}
\multiput(317,436)(10.298,-18.021){0}{\usebox{\plotpoint}}
\put(326.96,418.82){\usebox{\plotpoint}}
\put(337.83,401.14){\usebox{\plotpoint}}
\multiput(341,396)(11.513,-17.270){0}{\usebox{\plotpoint}}
\put(349.17,383.76){\usebox{\plotpoint}}
\put(361.38,366.98){\usebox{\plotpoint}}
\multiput(365,362)(12.966,-16.207){0}{\usebox{\plotpoint}}
\put(374.19,350.66){\usebox{\plotpoint}}
\put(387.98,335.15){\usebox{\plotpoint}}
\multiput(389,334)(13.789,-15.513){0}{\usebox{\plotpoint}}
\put(402.08,319.92){\usebox{\plotpoint}}
\multiput(405,317)(14.676,-14.676){0}{\usebox{\plotpoint}}
\put(417.00,305.50){\usebox{\plotpoint}}
\multiput(421,302)(15.620,-13.668){0}{\usebox{\plotpoint}}
\put(432.62,291.84){\usebox{\plotpoint}}
\multiput(437,288)(16.604,-12.453){0}{\usebox{\plotpoint}}
\put(448.94,279.04){\usebox{\plotpoint}}
\multiput(453,276)(17.270,-11.513){0}{\usebox{\plotpoint}}
\put(466.13,267.42){\usebox{\plotpoint}}
\multiput(470,265)(17.601,-11.000){0}{\usebox{\plotpoint}}
\put(483.73,256.42){\usebox{\plotpoint}}
\multiput(486,255)(17.601,-11.000){0}{\usebox{\plotpoint}}
\put(501.73,246.13){\usebox{\plotpoint}}
\multiput(502,246)(18.564,-9.282){0}{\usebox{\plotpoint}}
\multiput(510,242)(18.564,-9.282){0}{\usebox{\plotpoint}}
\put(520.30,236.85){\usebox{\plotpoint}}
\multiput(526,234)(18.564,-9.282){0}{\usebox{\plotpoint}}
\put(539.09,228.09){\usebox{\plotpoint}}
\multiput(542,227)(18.564,-9.282){0}{\usebox{\plotpoint}}
\multiput(550,223)(19.434,-7.288){0}{\usebox{\plotpoint}}
\put(558.15,219.95){\usebox{\plotpoint}}
\multiput(566,217)(19.434,-7.288){0}{\usebox{\plotpoint}}
\put(577.58,212.66){\usebox{\plotpoint}}
\multiput(582,211)(20.136,-5.034){0}{\usebox{\plotpoint}}
\put(597.29,206.27){\usebox{\plotpoint}}
\multiput(598,206)(20.136,-5.034){0}{\usebox{\plotpoint}}
\multiput(606,204)(19.434,-7.288){0}{\usebox{\plotpoint}}
\put(617.13,200.30){\usebox{\plotpoint}}
\multiput(623,199)(20.136,-5.034){0}{\usebox{\plotpoint}}
\put(637.31,195.42){\usebox{\plotpoint}}
\multiput(639,195)(20.136,-5.034){0}{\usebox{\plotpoint}}
\multiput(647,193)(20.136,-5.034){0}{\usebox{\plotpoint}}
\put(657.44,190.39){\usebox{\plotpoint}}
\multiput(663,189)(20.595,-2.574){0}{\usebox{\plotpoint}}
\put(677.76,186.31){\usebox{\plotpoint}}
\multiput(679,186)(20.136,-5.034){0}{\usebox{\plotpoint}}
\multiput(687,184)(20.595,-2.574){0}{\usebox{\plotpoint}}
\put(698.07,182.23){\usebox{\plotpoint}}
\multiput(703,181)(20.595,-2.574){0}{\usebox{\plotpoint}}
\put(718.38,178.15){\usebox{\plotpoint}}
\multiput(719,178)(20.595,-2.574){0}{\usebox{\plotpoint}}
\multiput(727,177)(20.595,-2.574){0}{\usebox{\plotpoint}}
\put(738.88,175.03){\usebox{\plotpoint}}
\multiput(743,174)(20.595,-2.574){0}{\usebox{\plotpoint}}
\multiput(751,173)(20.595,-2.574){0}{\usebox{\plotpoint}}
\put(759.38,171.95){\usebox{\plotpoint}}
\multiput(767,171)(20.595,-2.574){0}{\usebox{\plotpoint}}
\put(779.98,169.45){\usebox{\plotpoint}}
\multiput(784,169)(20.595,-2.574){0}{\usebox{\plotpoint}}
\multiput(792,168)(20.595,-2.574){0}{\usebox{\plotpoint}}
\put(800.58,166.93){\usebox{\plotpoint}}
\multiput(808,166)(20.595,-2.574){0}{\usebox{\plotpoint}}
\put(821.18,164.35){\usebox{\plotpoint}}
\multiput(824,164)(20.595,-2.574){0}{\usebox{\plotpoint}}
\multiput(832,163)(20.595,-2.574){0}{\usebox{\plotpoint}}
\put(841.77,161.78){\usebox{\plotpoint}}
\multiput(848,161)(20.756,0.000){0}{\usebox{\plotpoint}}
\put(862.43,160.20){\usebox{\plotpoint}}
\multiput(864,160)(20.595,-2.574){0}{\usebox{\plotpoint}}
\multiput(872,159)(20.595,-2.574){0}{\usebox{\plotpoint}}
\put(883.05,158.00){\usebox{\plotpoint}}
\multiput(888,158)(20.595,-2.574){0}{\usebox{\plotpoint}}
\put(903.68,156.04){\usebox{\plotpoint}}
\multiput(904,156)(20.756,0.000){0}{\usebox{\plotpoint}}
\multiput(912,156)(20.595,-2.574){0}{\usebox{\plotpoint}}
\put(924.37,155.00){\usebox{\plotpoint}}
\multiput(928,155)(20.595,-2.574){0}{\usebox{\plotpoint}}
\multiput(936,154)(20.629,-2.292){0}{\usebox{\plotpoint}}
\put(945.01,153.00){\usebox{\plotpoint}}
\multiput(953,153)(20.595,-2.574){0}{\usebox{\plotpoint}}
\put(965.70,152.00){\usebox{\plotpoint}}
\multiput(969,152)(20.595,-2.574){0}{\usebox{\plotpoint}}
\multiput(977,151)(20.756,0.000){0}{\usebox{\plotpoint}}
\put(986.39,150.83){\usebox{\plotpoint}}
\multiput(993,150)(20.756,0.000){0}{\usebox{\plotpoint}}
\put(1007.04,149.24){\usebox{\plotpoint}}
\multiput(1009,149)(20.756,0.000){0}{\usebox{\plotpoint}}
\put(1017,149){\usebox{\plotpoint}}
\sbox{\plotpoint}{\rule[-0.400pt]{0.800pt}{0.800pt}}%
\put(887,677){\makebox(0,0)[r]{Shannon}}
\put(909.0,677.0){\rule[-0.400pt]{15.899pt}{0.800pt}}
\put(220,344){\usebox{\plotpoint}}
\put(220.0,344.0){\rule[-0.400pt]{191.997pt}{0.800pt}}
\end{picture}

\end{figure}

\end{document}